\documentclass[prc,twocolumn,nofootinbib,superscriptaddress,showpacs]{revtex4}
\usepackage{graphicx,amsmath,amssymb,bm,multirow}
\usepackage{amsmath}
\usepackage{color}
\usepackage{amscd}

\begin{document}

\title{Pairing correlations of cold fermionic gases at overflow from a narrow to a wide harmonic trap}

\author{A. Pastore} \email{apastore@ulb.ac.be} 
\affiliation{Institut d'Astronomie et d'Astrophysique, CP 226,
  Universit\'e Libre de Bruxelles, B-1050 Bruxelles, Belgium}
\author{P. Schuck} 
\affiliation{Institut de Physique Nucl\'eaire, Universit\'e Paris-Sud,
  IN2P3-CNRS, F-91406 Orsay, France}
\affiliation{  Laboratoire de Physique et de Mod\'elisation des Milieux Condens\'es, CNRS et Universit\'e Joseph Fourier, UMR5493, 25 Av. des Martyrs, BP 166, F-38042 Grenoble Cedex 9, France}
\author{M. Urban} 
\affiliation{Institut de Physique Nucl\'eaire, Universit\'e Paris-Sud,
  IN2P3-CNRS, F-91406 Orsay, France}
\author{X. Vi\~nas} 
\affiliation{Departament de structura i Constituents de la Mat\`eria
  and Institut de Ci\`encies del Cosmos, Facultat de Fisica,
  Universitat de Barcelona, Diagonal 647, E-08028 Barcelona, Spain}
\author{J. Margueron} 
\affiliation{Institut de Physique Nucl\'eaire, Universit\'e Paris-Sud,
  IN2P3-CNRS, F-91406 Orsay, France}
\affiliation{Universit\'e de Lyon, Institut de Physique Nucl\'eaire de
  Lyon, IN2P3-CNRS, F-69622 Villeurbanne, France}

\date{\today}

\begin{abstract}
Within the context of Hartree-Fock-Bogoliubov theory, we study the
behavior of  superfluid Fermi systems when they pass from a small to a
large container. Such systems can be now realized thanks to recent
progress in experimental techniques. It will allow to better
understand pairing properties at overflow and in general in rapidly varying external potentials.
\end{abstract}


\maketitle


\section{Introduction}\label{sec:intro}

Finite systems of superfluid fermions exist in different situations
such as, for example, traps of cold atoms, atomic nuclei, or small
metallic clusters.  All these different systems can be treated using
the same many-body theory, namely the Bogoliubov-de Gennes (BdG)
theory~\cite{bar57,DeGennes1999}, in nuclear physics usually called
Hartree-Fock-Bogoliubov (HFB) theory, valid also for finite range 
pairing forces~\cite{Book:Ring1980}.  In a recent
series of articles~\cite{Schuck2011,Schuck2012}, the authors have
studied the behavior of different systems of fermions once their Fermi
levels reach the edge of a finite container, $ i.e.$, either the
fluid overflows into the continuum or it pours into another container
of much larger dimension.  Such a situation has been experimentally
observed in systems of cold bosonic atoms \cite{StamperKurn1998,Web03}
 and it should also be possible to use it for fermionic
atoms \cite{Viverit2001}.  This phenomenon is supposed to be realized also 
in the
inner crust of neutron stars (NS) in the transition region between the
outer and the inner crust~\cite{Chamel2008a}.  Going from the outer
layer to the inner one, $i.e.$ from low to high density regions, the
formation of very neutron rich nuclei is favored by the NS environment
and eventually they can reach the neutron drip line.  Further
increasing the density, the neutrons can not be bound anymore and we
observe the coexistence of a lattice of nuclei immersed in a diluted
neutron gas. This situation often is mimicked by the use of
Wigner-Seitz (WS) cell~\cite{Wigner1934a,Chamel2007}, where the single
particle potential has a pocket, representing the nucleus, embedded in
a large container.  Understanding the role of pairing correlations in
these systems is thus important and will help us to get general insight into 
the behavior of superfluid Fermi systems in rapidly varying external 
potentials.

The aim of the present article is to give considerably more details and to 
extend the study of ~\cite{Schuck2011,Schuck2012}, considering cold atoms 
as a relatively simple bench mark system. In cold atom systems the 
interparticle distance is by orders of magnitude larger than the range of the 
force between atoms, so that the latter is usually replaced by a contact 
interaction with a well-known regularisation 
scheme~\cite{yu02,Grasso2003}. 
The systems considered in ~\cite{Schuck2011,Schuck2012} were treated 
not with the BdG (HFB) theory but using the BCS approximation in which 
non-diagonal matrix elements of the gap are neglected. These 
calculations were performed using a bare contact force with a cut-off.
However, the BCS approximation fails in the case of cold atoms  
when a contact force with the usual regularisation procedure  is 
used. We have checked this numerically but it 
may also become 
clear from the fact that the BCS wave function, contrarily to the HFB-one, 
has not the correct asymptotic behavior when the relative distance between 
the two particles goes to zero. It is, therefore, an important issue to see 
by how much the conclusions of ~\cite{Schuck2011,Schuck2012} are altered 
in using the correct BdG(HFB) approach. We will see that qualitatively the 
quench of the gap at overflow persists.  However, we will have to correct 
one conclusion drawn from the BCS approximation concerning the spatial 
behavior of the gap in the interior of the trap where the BCS 
solution does not behave as the BdG(HFB) one.
We also will present new features as, for instance, the spatial behavior 
of the wave functions before and after the spill which will give more insight 
into the fact why the gap becomes quenched passing the drip.

%

The article in organized as follows: in Sec.\ref{Method} we present
the HFB equations we used in our calculations and we also discuss the
properties of the adopted pairing interaction. In Sec.\ref{sec:res},
we present our numerical results and finally in Sec.\ref{sec:concl},
we give our conclusions.



\section{The Hartree-Fock Bogoliubov or Bogoliubov-de Gennes Approach}\label{Method}

To discuss the pairing properties of a system of cold atoms, we adopt
the mean-field description based on the HFB
equations~\cite{Book:Ring1980} or, equivalently, the BdG
equations~\cite{DeGennes1999}. The difference between both methods is
marginal: the BdG equations are taylored for zero range forces, {\it
  i.e.}  local pairing fields, whereas nuclear forces are in general
finite ranged and, thus, also the pairing field becomes non-local
demanding the more general HFB scheme. Using an effective zero range
interaction, we can express the HFB (BdG) equations in $r$-space
as~\cite{Grasso2003,bruun1999bcs}
\begin{eqnarray}\label{HFB:eq}
[H_{0}+W(\mathbf{R})]u_{\alpha}(\mathbf{R})+\Delta(\mathbf{R})v_{\alpha}(\mathbf{R})=E_{\alpha}u_{\alpha}(\mathbf{R})\nonumber\\
\Delta(\mathbf{R})u_{\alpha}(\mathbf{R})- [H_{0}+W(\mathbf{R})]v_{\alpha}(\mathbf{R})=E_{\alpha}v_{\alpha}(\mathbf{R}).
\end{eqnarray}
We have used $\alpha=\{nls\}$ as a short-hand notation for the quantum
numbers of the system; since we have chosen a system with perfect
symmetry between states of opposite spin
$N/2=N_{\uparrow}=N_{\downarrow}$, for brevity, we can simply drop the
spin quantum number $s$. In Eqs.(\ref{HFB:eq}), $H_{0}=T+U(R)-\mu$; $T
= -\hbar^2\nabla^2/2m$ is the kinetic energy, $\mu$ is the chemical
potential and $U(R)$ is the external trap potential. $W(\mathbf{R})$
is the Hartree potential. We will neglect it in this paper, since it
is not essential for our following considerations. Let us again point
out that the present work may have generic character for all
situations where a superfluid Fermi liquid pours from a narrow to a
much wider container. We will consider quite large systems where shell effects are not very important, but  we briefly will also touch smaller ones. 
As  it is known, in the case of nuclei 
the superfluid mechanism can be strongly influenced by the
underlying level structure~\cite{mar12,Pastore2013}

\begin{figure}
\begin{center}
\includegraphics[clip=,width=0.35\textwidth,angle=-90]{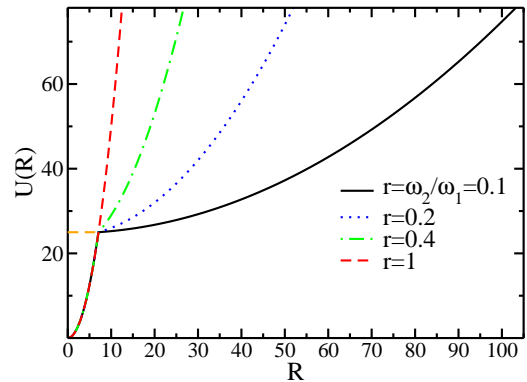}
\end{center}
\caption{(Colors online) The different trapping potentials $U(R)$ used
  in this work, as defined in Eq.(\ref{eq:u:trap}). The ratio among
  the two harmonic potentials is $r=\omega_{1}/\omega_{2}=0.1,0.2,0.4,1$. $U$ and $R$ are in trap units, \emph{i.e.}, in units of $\hbar\omega_{1}$ and $\sqrt{\hbar/m\omega_{1}}$, respectively [cf. text below Eq.(\ref{eq:u:trap})]. 
}
\label{utrap}
\end{figure}

In the present article, following the discussion of
Refs.~\cite{Schuck2011,Schuck2012} we have adopted a spherical
double-harmonic potential of the form
\begin{eqnarray}\label{eq:u:trap}
U(R)=\left\{ \begin{array}{cc} \frac{1}{2} m\omega_{1}^{2} R^{2} &
  \text{if } R<R_{0}\,, \\ 
\frac{1}{2} m[\omega_{2}^{2} R^{2}+(\omega_{1}^{2}-\omega_2^2) R_{0}^{2}]& \text{elsewhere.}
\end{array}\right.
\end{eqnarray}
From now on, we will use the system of units of the trap,
\textit{i.e.}, $\hbar=\omega_1=m=1$. In other words, energies are
measured in units of $\hbar\omega_1$, lengths in units of the harmonic
oscillator length $\sqrt{\hbar/m\omega_1}$, etc. We choose $R_0 =
\sqrt{50} \approx 7.1$ in these units. In Fig.\ref{utrap}, we show the
radial profile of the external trap. We observe that according to the
ratio $r=\omega_{2}/\omega_{1}$, we can define different
configurations. Going from $r=1$ to $r=0.1$, we create an overflow
point at $U(R_{0})$=25.

The HFB equations~(\ref{HFB:eq}) are solved in the basis of the
eigenfunctions of the potential~(\ref{eq:u:trap}) up to a maximum
energy $E_C$. A similar method has been already described in
ref~\cite{Pastore2009} for the nuclear case.

The pairing potential $\Delta(R)$ originates from a contact pairing
interaction. As already discussed in Ref.~\cite{Bulgac2002}, the
contact interaction presents an ultraviolet divergence. To avoid such
a problem we can either fix the phase space where we perform our
calculations by introducing a
cut-off~\cite{Schuck2011,Schuck2012,Borycki2006} or we can identify
the nature of the divergent term and regularize the interaction. The
latter procedure has been suggested in Ref.~\cite{Bulgac2002} and it
will be the method followed in the present article.

Following Ref.~\cite{Grasso2003}, we write the gap equation with an
effective contact pairing interaction as
\begin{eqnarray}\label{delta:eff}
 \Delta(R)=g_{\text{eff}}(r)\sum_{nl}\frac{2l+1}{4\pi}u_{nl}(R)v_{nl}(R)\,.
 \end{eqnarray}
Since we work in a finite basis, $n$ and $l$ cannot exceed maximum
values defined by the cut-off energy $E_C$. However, the results will
not depend on the choice of $E_C$ if it is sufficiently large. This
cut-off should not be confused with the one adopted in
Refs.~\cite{Schuck2011,Schuck2012}, since in that case the authors
have used a bare contact force $g$, while here we use an effective
contact force $g_{\text{eff}}$ which compensates the cut-off
dependence. This effective interaction reads
\begin{eqnarray}\label{geff}
\frac{1}{g_{\text{eff}}}=\frac{1}{g}+\frac{1}{2\pi^{2}}\left[\frac{k_{F}(R)}{2}\log
  \frac{k_{C}(R)+k_{F}(R)}{k_{C}(R)-k_{F}(R)} -k_{C}(R)\right]\,,
\end{eqnarray}
where $k_{F}(R)=\sqrt{2(\mu-U(R))}$ is the local Fermi momentum and
$k_{C}(R)=\sqrt{2(E_{C}-U(R))}$ is the momentum corresponding to the cut-off.

The coupling constant $g$ is related to the atom-atom scattering
length $a$ by $g = 4\pi a$. Unless otherwise stated, we will use $g=-1.56$, corresponding to $k_F a = -0.88$ at the center of the trap
for $\mu = 25$. Those numbers fall well into the domain what is attainable with the help of Feshbach resonances in real cold atom systems. Our study will, therefore, provide a definite experimental prediction.

As already discussed in ref. \cite{Grasso2003}, we have to check that
our calculations are fully converged, {\it i.e.}, that they do not
depend on the given value of $E_{C}$.

\begin{figure}
\begin{center}
\includegraphics[clip=,width=0.35\textwidth,angle=-90]{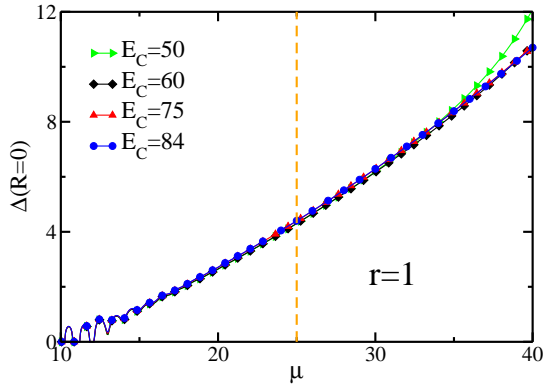}
\end{center}
\caption{(Colors online) Evolution of the pairing field at the center of the trap $\Delta(R=0)$ as a function of the chemical potential  $\mu$ in the trap configuration $r=\omega_{2}/\omega_{1}=1$ for different cut-off values $E_{C}$  for g=-1.56. All quantities are expressed in trap units  [cf. text below Eq.(\ref{eq:u:trap})].}
\label{error}
\end{figure}

In Fig.\ref{error}, we show for the trap configuration  the evolution of the 
pairing field at the center of the trap $\Delta(R=0)$ as a function of the 
chemical potential and different values of the cut-off $E_{C}$.
The calculations are well converged for the choice $E_{C}=84$  around small 
values of the chemical potential, and in particular for the region around 
$\mu=25$, where we have the drip point. See Fig.\ref{utrap}.
The result is in agreement with the conclusions of Ref.~\cite{Grasso2003}. 
At higher values of the chemical potential, we notice that the adopted value 
of $E_{C}$ is not very adequate and a  larger value would be 
preferable. 
As already mentioned in the introduction, we again want to stress here that 
the outlined theoretical scheme can in no way be approximated by the simple 
BCS approach of \cite{Schuck2011,Schuck2012} where the off-diagonal elements 
of the gap are neglected.\\

\begin{figure}
\begin{center}
\includegraphics[clip=,width=0.35\textwidth,angle=-90]{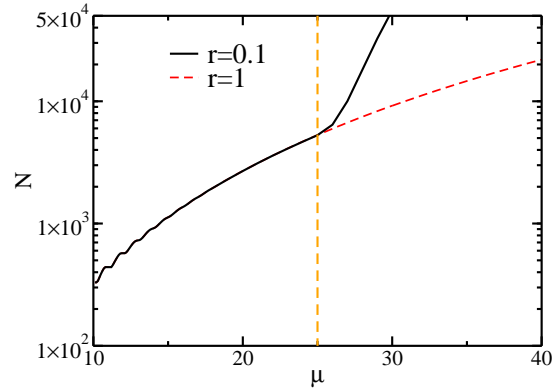}
\end{center}
\caption{(Colors online) Number of particles as a function of the chemical potential  (in units of $\hbar\omega_1$)  for two trap configurations.}
\label{nparticle}
\end{figure}

In order to show the typical system sizes we are dealing with in this work, we display in Fig.\ref{nparticle} the number of atoms in the trap as a function of the chemical potential $\mu$ for two trap configurations.  Please, notice the practically exponential increase of particle numbers, once the large container starts being populated.

\section{Results}\label{sec:res}

Having checked the numerical accuracy that we can achieve
in our calculation, we can finally present our results.

\subsection{HFB}

\begin{figure*}
\begin{center}
\includegraphics[clip=,width=0.35\textwidth,angle=-90]{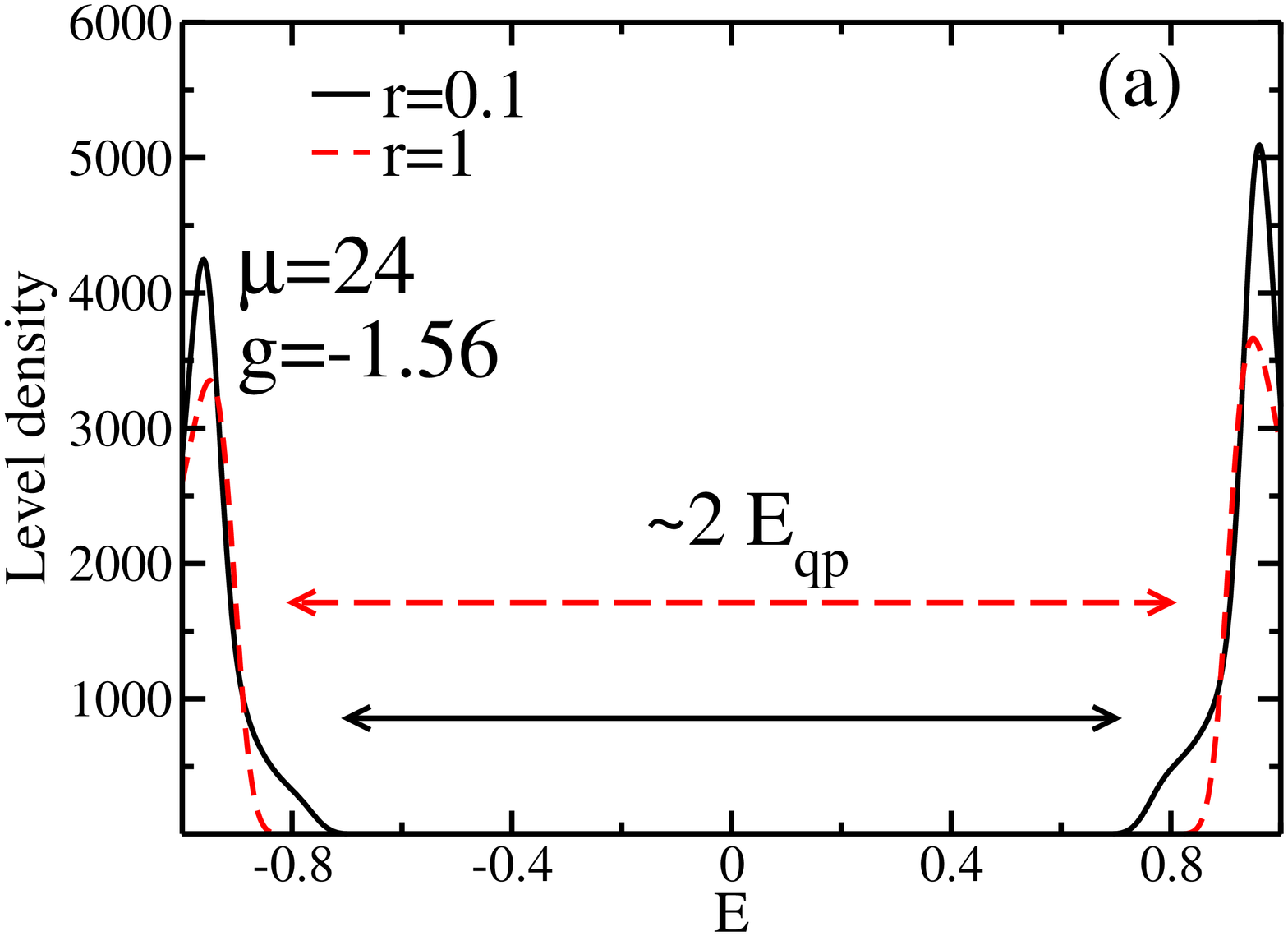}
\hspace{-2.38cm}
\includegraphics[clip=,width=0.35\textwidth,angle=-90]{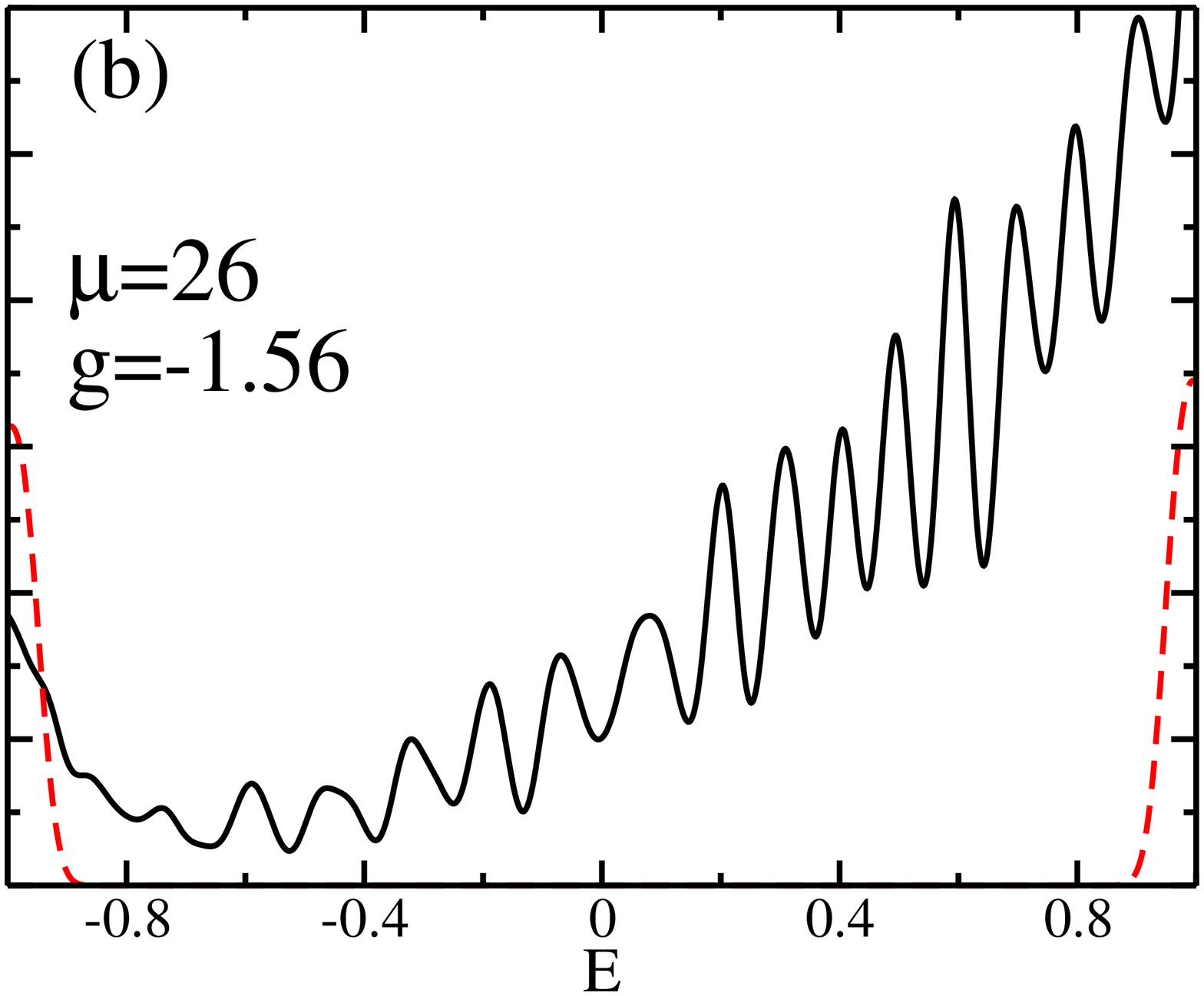}\\
\vspace{-0.9cm}
\includegraphics[clip=,width=0.35\textwidth,angle=-90]{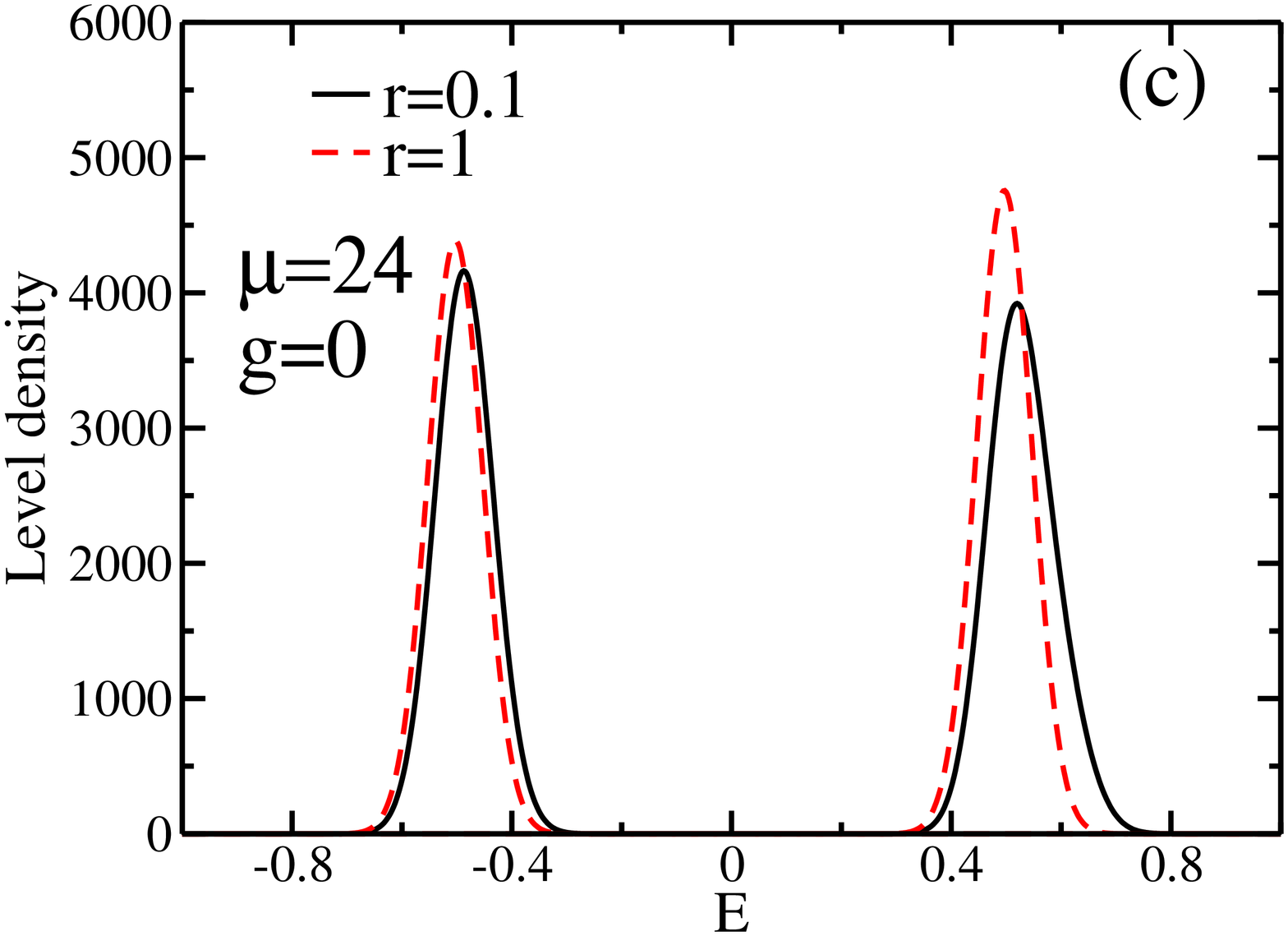}
\hspace{-2.38cm}
\includegraphics[clip=,width=0.35\textwidth,angle=-90]{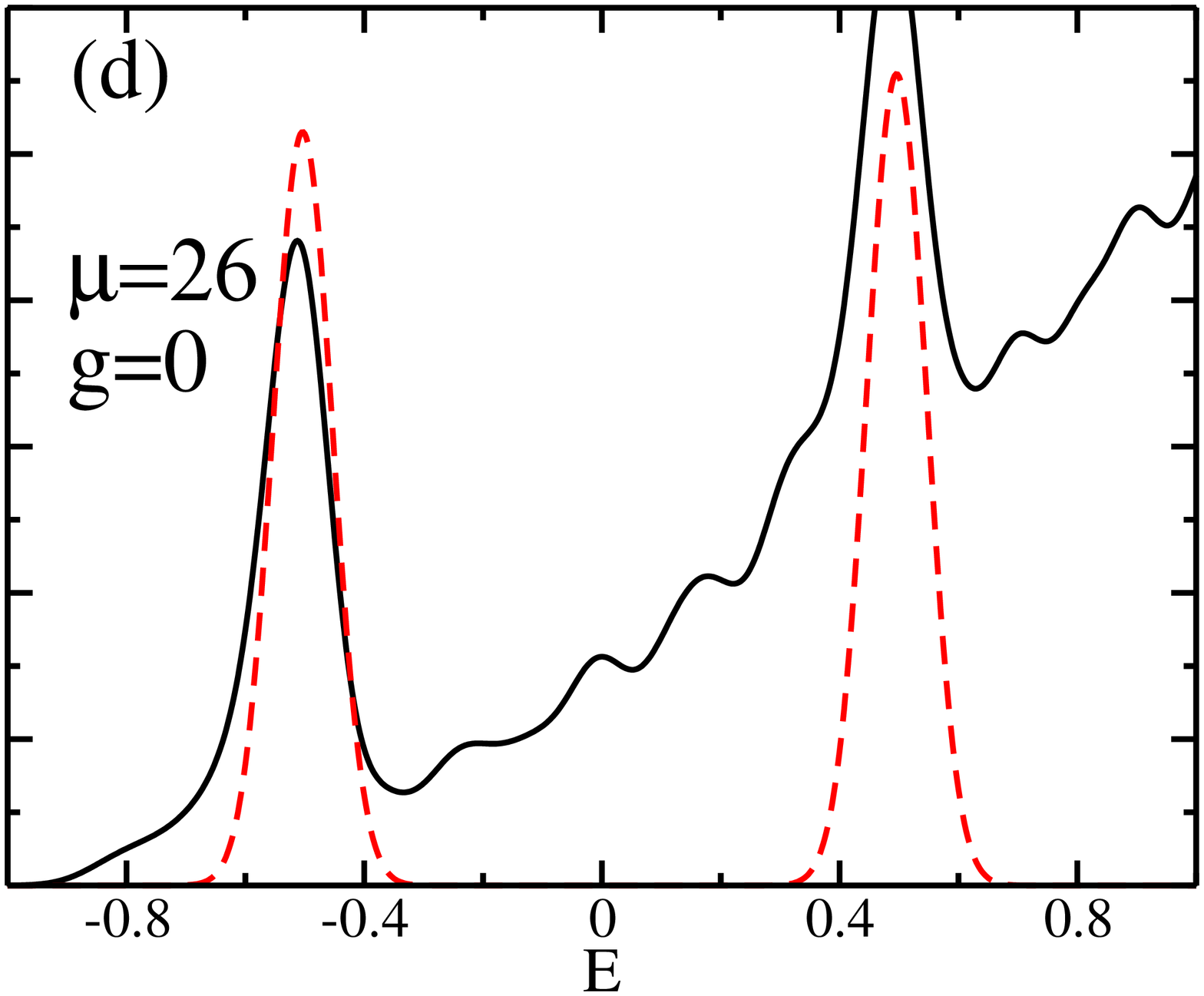}
\end{center}
\caption{(Color online)  Level density as given in Eq.(\ref{eq:lev:dens}) for
  the two extreme traps as shown in Fig.\ref{utrap}. On the left
  panels the calculations are done at $\mu=24$, while on the right panels
  $\mu=26$. In panels (a)-(b) we show the level densities from  HFB calculations and in (c)-(d) the ones without pairing. The discrete level densities are
  folded with gaussians of width $\sigma=0.05$ for the case with
  pairing (panels a and b) and $\sigma=0.1$ for the case without
  pairing (panels c and d). All quantities are expressed in trap units.
 }
\label{lev:dens}
\end{figure*}

\begin{figure}
\begin{center}
\includegraphics[clip=,width=0.36\textwidth,angle=-90]{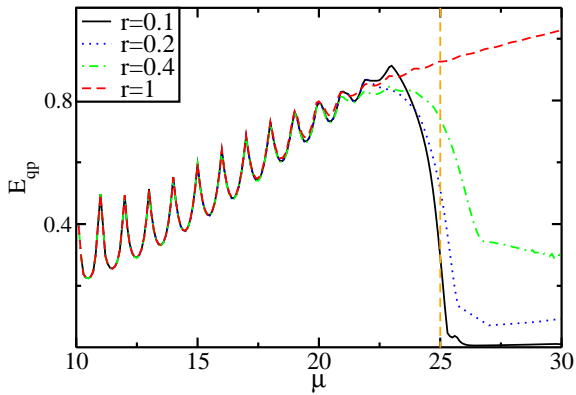}
\end{center}
\caption{(Colors online) Evolution of the lowest quasi-particle energies as a function of the chemical potential (both in units of $\hbar\omega_1$) for various potential configurations.}
\label{eqp:min}
\end{figure}

One of the most prominent indicators of superfluidity is the presence
of a gap in the level density. It may not be very easy to measure it
in cold atom systems, but in metallic superconductors the level
density around the Fermi level is measured routinely~\cite{ral95}. May
be, if one day a metallic grain device can be constructed with small
and wide container geometry as discussed here, it may be possible to
measure the level density across the drip point. In any case, we will
discuss the result of our calculation for the superfluid level
density. A simplified version of the level density for the BCS case
has been also presented in ref.~\cite{Schuck199612}. Here, we give the
full expression
\begin{multline}\label{eq:lev:dens}
g(E)=2\sum_{E_{nl}>0}(2l+1)\left[ \|v_{nl}\|^2\delta\left(E+E_{nl}\right)\right.\\
  +\left. \|u_{nl}\|^2\delta\left( E-E_{nl}\right)\right]
\end{multline}
where $E$ is the energy measured from the chemical potential $\mu$,
$E_{nl}$ is the HFB quasi-particle (qp) energy, and $\|v_{nl}\|$,
$\|u_{nl}\|$ denote the norms of the wave functions $v_{nl}$ and
$u_{nl}$, respectively. Results for various system parameters are
shown in Fig. \ref{lev:dens}. In Fig.\ref{lev:dens}(a)-(b) we show the
level density  for $\mu$-values slightly below and slightly above the drip point,
respectively. For the ratio $r=0.1$, we clearly see a break down of
the gapped window when passing from below to above the drip. Also
shown is the level density without pairing for better comparison
(panels \ref{lev:dens}(c)-(d)).  Actually it is easier to represent
the lowest qp energy as a function of the chemical potential, since
the gap in the spectrum is essentially given by two times the lowest
qp energy.  This is shown in Fig. \ref{eqp:min}. We clearly see that
there is a strong break down of the gap passing through the drip point
at $\mu = 25$. This drop is the more pronounced, the smaller the ratio
$r$. For $r=0.1$, the gap drops practically to zero after the drip. It
is interesting to note that the decrease of the gap actually starts
already {\it before} reaching the drip point and precisely at overflow the gap is already down by an important factor. This result is quite
well in agreement with the qualitative prediction given in
\cite{Schuck2011} where, however, the more crude BCS approximation was
used.   We also can see from Fig. \ref{eqp:min} that once the chemical potential is well below the drip point, quite prominent shell fluctuations start to develope. Shell fluctuations will also appear in other quantities and further be discussed in section~\ref{sec:LDA}. It is rather easy to understand why there is this drop of the
spectral gap and eventually other pairing quantities across the
overflow point. Since the individual wave functions after the
threshold have a much wider extension, still being normalized to
unity, their overlap in the matrix elements of the pairing force
becomes much smaller. What, nevertheless, is surprising, is the
sharpness of the transition displayed in Fig. \ref{eqp:min}.
 Numerically, we cannot handle still smaller values of $r$ than $r=0.1$. However, from Fig.\ref{eqp:min}, it becomes quite clear that in the limit $r \rightarrow 0$, \emph{i.e.}, for a truly finite potential, the lowest quasiparticle and, thus, the gap will become very close to zero, once the chemical potential $\mu$ reaches the top of the finite container.

\begin{figure}
\begin{center}
\includegraphics[clip=,width=0.36\textwidth,angle=-90]{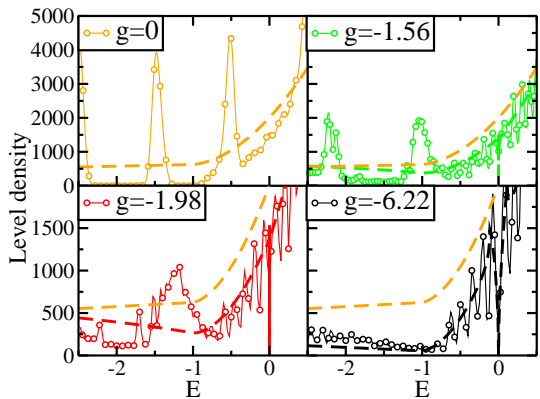}
\end{center}
\caption{(Colors online) Level density as given in
  Eq.(\ref{eq:lev:dens}) for the trap configuration
  $r=\omega_{2}/\omega_{1}=0.1$ at $\mu=26$ and four different values of
  coupling constant. On the same figure we represent with dashed line
  and same color code the results obtained using the LDA
  approximation. All quantities are expressed in trap units.}
\label{lev:dens:diff:g}
\end{figure}

In order to make the influence of pairing more pronounced, we repeated
the calculation using different values of the coupling constant $g$. In
Fig. \ref{lev:dens:diff:g}, we present a zoom on the level density in
the energy interval $E\in[-2.5, +0.5]$ (where the energy is measured
from the chemical potential $\mu$) for four values of the coupling
constant $g$ as indicated in the panels. In each panel, we also give
the level density in LDA (broken lines), see Sect.\ref{sec:LDA}. The
underlying basis is the same for the four different calculations and
we just change the value of the coupling constant $g$. We notice that
for a range of coupling constants $g\lesssim2$, the pairing
correlations are not strong enough to create a gap in the level
density on the scale of the figure. Only, once we increase the
coupling constant by a factor of four, we can observe the appearance
of a tiny gap in the LDA level density around $E=0$. Quantally ({\it
  i.e.}, within HFB), there are only discrete levels and it is very
difficult to pin down any opening of a gap in this case. For
comparison, we have also represented the case without pairing obtained
with LDA in all panels (yellow broken lines). We not only see the
(narrow) window at $E=0$, but remark that even globally the level
density is suppressed in the superfluid case as seen for instance for
 $g=-6.22$ (lower right panel of Fig.\ref{lev:dens:diff:g}). This stems from the fact that
for this increased pairing force, the gap in the narrow part of the
potential is so strong that it still influences the level density far
from the Fermi energy ({\it i.e.}, far from $E=0$).

\begin{figure}
\begin{center}
\includegraphics[clip=,width=0.38\textwidth,angle=-90]{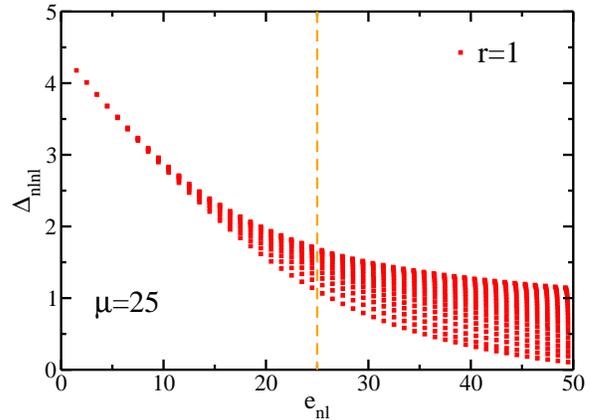}\\
\includegraphics[clip=,width=0.38\textwidth,angle=-90]{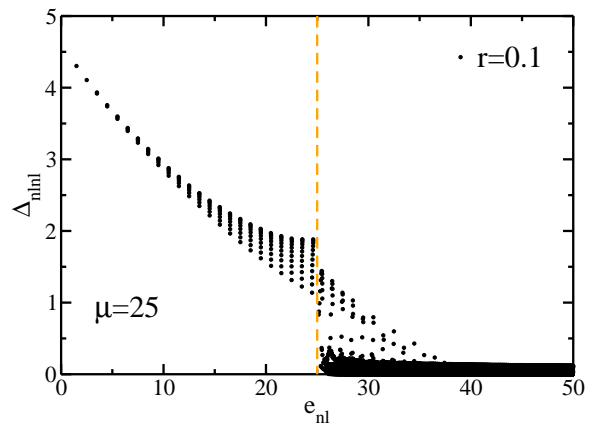}
\end{center}
\caption{(Colors online) Diagonal matrix elements of the pairing
  field, $\Delta_{nlnl}$, as a function of the energy of the basis
  state $n$ at $\mu=25$ for two different configurations  ($\Delta_{nlnl}$, $e_{nl}$, and $\mu$ in units of $\hbar\omega_1$).  In the upper panel we use the trap configuration $r=1$, while in lower panel $r=0.1$.}
\label{mel:ef50}
\end{figure}

We have also analyzed the diagonal matrix elements of the gap (as they are used in the BCS approximation, see Refs.~\cite{Schuck2011,Schuck2012} ) as a
function of the single particle energies as they are given in our
basis, see Fig.~\ref{mel:ef50}. We observe a clear drop of most of those
gap values beyond the overflow point at $\mu = 25$ in the case $r=0.1$. However, what is
interesting is the fact that some of those gaps in the region $\mu > 25$
apparently still belong to the narrow container. As already discussed
in the nuclear case \cite{mar12,Pastore2013}, those gaps correspond to
'resonances' of the narrow container in the 'continuum' (actually
levels are, of course, still discrete above the drip point but very
dense, \emph{i.e.} one may consider it as a quasi-continuum). Those
resonances correspond in the nuclear case to states trapped by the
centrifugal barrier of the narrow container. We suppose that the same
features are at work in our present case.

\begin{figure}
\begin{center}
\includegraphics[clip=,width=0.35\textwidth,angle=-90]{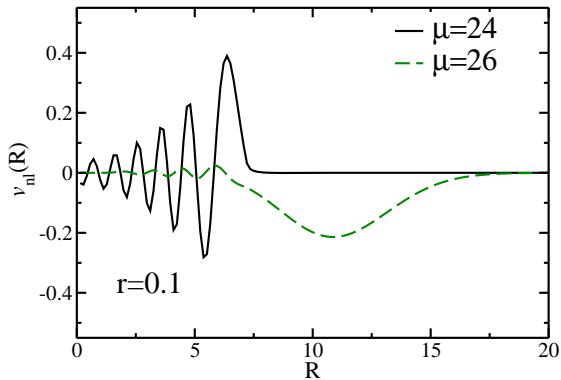}\\
\end{center}
\caption{(Colors online) The radial spinor $v_{nl}(R)$ as defined in
  Eq.\ref{HFB:eq} corresponding to the lowest quasi-particle as shown
  in Fig.\ref{eqp:min} for the trap configuration $r=0.1$. The solid
  line refers to the case $\mu=24$, while the dashed line to the case
  $\mu=26$. All quantities are expressed in trap units.}
\label{psi:min:g11}
\end{figure}

It is also interesting to study the behavior of the wave functions
$v(R)$ ($u(R)$ is similar). 
In Fig.~\ref{psi:min:g11},
we show
$v_{nl}(R)$  corresponding to the lowest quasi-particle energy for the case of $r = \omega_2/\omega_1 = 0.1$ for two
values of the chemical potential $\mu = 24$ and $\mu = 26$. We see that the behavior of the wave
function for the case of $\mu = 24$ is quite similar to the case presented by
Baranov~\cite{baranov} and by Bruun and Heiselberg~\cite{bruun2002cooper}, namely that the wave function is suppressed in the center of the trap (where $\Delta(R)$ is large) and is mostly concentrated on the surface of the gas.. 
 
The situation changes quite significantly in the case of $\mu = 26$.
The extension of
the wave function is radically increased and spills far into the wide
container with a large amplitude. Such an effect is, of course,
expected but it is of interest to see it quantitatively. This
behavior explains the small energy of the lowest quasiparticle energy
after the overflow shown in Fig.~\ref{eqp:min}.



\subsection{LDA}\label{sec:LDA}

\begin{figure}
\begin{center}
\includegraphics[clip=,width=0.3\textwidth,angle=-90]{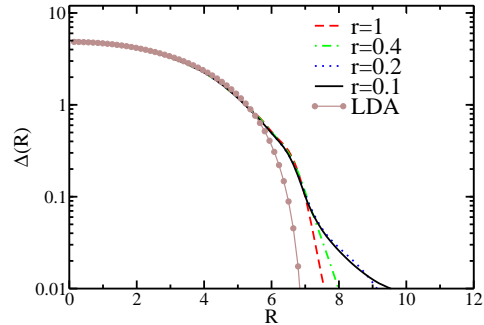}\\
\end{center}
\caption{(Colors online)Pairing field $\Delta(R)$ calculated at $\mu=26$ for different trap configurations. The filled circles indicate the result obtained using the LDA for $r=0.1$. All quantities are expressed in trap units.}
\label{deltar:ef52}
\end{figure}

\noindent
In cold atom systems, it is common use to solve the pairing problem in the homogeneous case and then to use
 the local-density approximation (LDA)~\cite{Book:Ring1980}, which
amounts to replacing $\Delta$ by $\Delta(R)$
and $\mu$ by $\mu(R) = \mu - U(R)$. Usually this works quite well as
has recently been discussed in detail in ref.~\cite{Grasso2003}. So,
let us see how LDA works in the present case. 

Again the interesting quantity may be the level density in LDA. How to
calculate the level density within LDA is demonstrated
in~\cite{Schuck199612} (but here we make use of the LDA version of Eq.(\ref{eq:lev:dens}) which is slightly more general than the one given in ~\cite{Schuck199612}). We show the results in
Fig.\ref{lev:dens:diff:g} (broken lines). In this figure we display
the LDA results using different values of the coupling constant $g$
and fixed trap configuration $r = \omega_{2}/\omega_{1}=0.1$.  The
results are in good agreement with the quantal ones in the sense that the LDA is able
to reproduce the average quantities thus passing through the peaks
obtained in the HFB calculations. We also observe, as already
mentioned, that the LDA level density always goes to zero at the Fermi
energy ({\it i.e.}, at $E=0$) as it should be, but on the scale of the
figure one only can notice a very small gap for the case of the
increased pairing force $g=-6.22$.
  
\begin{figure}
\begin{center}
\includegraphics[clip=,width=0.3\textwidth,angle=-90]{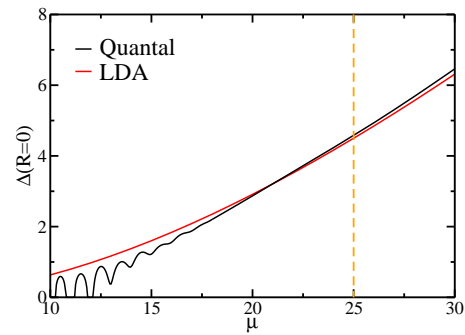}
\end{center}
\caption{(Colors online) Comparing the quantal and LDA calculations of
  pairing field $\Delta(R=0)$ as function of the chemical potential $\mu$ (both in units of $\hbar\omega_1$) for the trap configuration $r=0.1$.}
\label{lda:compare}
\end{figure}

In systems of cold atoms, one can measure the gap locally~\cite{sch09}. In Fig.~\ref{deltar:ef52}, we show the local gap
$\Delta(R) $ as a function of position for the quantal case
compared to LDA for the ratio $r=0.1$. It is seen
that at large radii, the LDA result drops too fast (as soon as
$\Delta(R)$ becomes smaller than the level spacing of the trap).
Quantitatively, for expectation values, this may not be very relevant,
however, because the logarithmic scale on the vertical axis in
Fig.~\ref{deltar:ef52} makes the effect look more important than it
is.

Let us now look at the maximum of the local gap, which is at the
origin of the trap. In Fig.\ref{lda:compare} we show $\Delta(R=0)$ as
a function of $\mu$ quantally and in LDA. We see that this quantity is
monotonously and constantly  rising, also across the drip point and
that quantal and LDA results well agree on the high energy side (the
very small disagreement likely is an effect of not 100\% converged
results as the chemical potential reaches higher values). This,
however, is not in contradiction with the fact that the spectral gap
is sharply dropping after the overflow as seen in Fig.\ref{eqp:min}, since 
the spectral gap depends mainly on what happens at the surface of the system,
 see Fig.\ref{psi:min:g11}.
An important qualitative difference between the BCS prediction and 
the BdG-HFB approach can be seen on Figs. \ref{deltar:ef52},\ref{lda:compare}
. Whereas at overflow the BCS approximation entails a strong negative 
influence on what happens for the gap also at the center of the trap, 
this is not at all the case with the BdG-HFB, see, for instance, 
Fig. \ref{deltar:ef52}. On the contrary, as seen on Fig. 5 of 
\cite{Schuck2011}, in the BCS case, the gap inside seems to disappear 
alongside with the disappearance of the gap outside, thus, invalidating 
the BCS approximation for this feature.

In principle other quantities may be of interest. For example how does
the difference of energy per particle, $\Delta E$ with and without
pairing vary as a function of the chemical potential? The same
question may be asked for the pairing energy, etc. However, the latter
quantity diverges (as a function of $E_C$) for the present contact
interaction, only the total energy converges. On the other hand the
(tiny) energy difference $\Delta E$ comes as a result of the
difference of two big numbers and the accuracy of our code did not
allow us to obtain totally stable results. Thus, we do not present
them here. However, our estimates are such that, \emph{e.g.}, the drop of
pairing correlations across the overflow, seen already for the level
density and the lowest quasi particle energy, is confirmed also for
the above mentioned difference of energies.\\

\subsection{Small particle number}
\begin{figure}
\begin{center}
\includegraphics[clip=,width=0.36\textwidth,angle=-90]{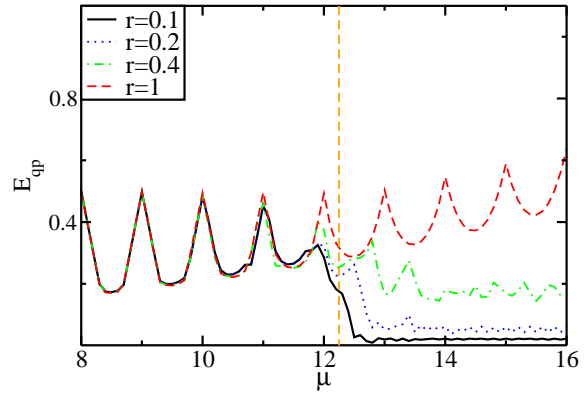}
\end{center}
\caption{(Colors online) Same as Fig.\ref{eqp:min}, but for overflow at $\mu = 12.25$ and $R_0 = 4.95$ (in trap units). See text for more explanation.}
\label{eqp:min2}
\end{figure}

\begin{figure}
\begin{center}
\includegraphics[clip=,width=0.36\textwidth,angle=-90]{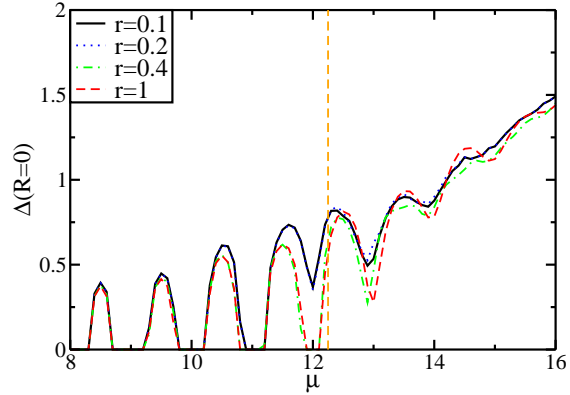}
\end{center}
\caption{(Colors online) Pairing field at the center of the trap $\Delta(R=0)$ as a function of the chemical potential $\mu$ (both in units of $\hbar\omega_1$) for the  same trap configuration as in Fig.\ref{eqp:min2} and different ratios $r$.}
\label{d0:smalltrap}
\end{figure}

 One may think that the features outlined above for the double harmonic potential may change if in the narrow part of the potential are not several thousands but only several hundreds  of particles reaching the drip, a situation which prevails, \emph{e.g.}, in the nuclear case. Note that also in traps it is now possible to study systems with very small atom numbers \cite{Serwane2011}.
We, therefore, show in Fig.\ref{eqp:min2} again the lowest quasi particle energies as a function of $\mu$ for a case where the overflow occurs at $\mu = 12.25$ corresponding to an extension of the narrow part of the trap $R_0 = 4.95$, see Fig.~\ref{utrap}. In Fig.\ref{d0:smalltrap} the gap at the center is displayed as a function of $\mu$. A little unexpectedly the situation stays qualitatively similar to the cases shown in Fig.\ref{eqp:min} and Fig.\ref{lda:compare}, only the influence of the shell effects is now much stronger.

\subsection{Physical Consequences}

As we mentioned already, in some systems the gap can directly be
measured via the level density. It remains an open question whether
for instance metallic grains can geometrically be tailored into such a
double well configuration. In other systems, like nuclei and ultra
small metallic grains the even-odd effect, directly related to the
gap, can be extracted from experimental data~\cite{Tinkham}. As we mentioned already, in
nuclei with their quite small number of particles, the drop of pairing
is strongly overshadowed by shell fluctuations. However, the tendency
of  reduced pairing correlations towards, for instance, the neutron
drip is also clearly seen there from theoretical
calculations~\cite{Pastore2013,mar12}.  However, static gaps being
interesting, the more spectacular expression of superconductivity and
superfluidity comes from the dynamics, as for example collective modes
or rotational motion, etc. One may  be interested in
situations where the linear response regime applies and, thus, only
the configurations very close to the Fermi level are active. So, since
the gap values drastically drop in the overflow regime, this will also
affect the dynamics. For example, because of the reduced gap, pair
breaking (at small but finite temperature) will occur much more
frequently after the drip into the large container than before and,
therefore, the collective flow will become more normal fluid like. To
be more quantitative, let us discuss the situation of slow rotation~\cite{far00,urb03}.  In this (and other)
situation(s) it is not so much the absolute value of the spectral gap
which plays a role but rather the ratio of the gap vs the average
level spacing. In our case, this would mean to scale the gap to the
frequencies $\omega_i$. $\Delta/\omega > 1$ is the regime of
irrotational flow, whereas $\Delta/\omega < 1$ is the regime of rigid
body motion. Of course around the drip it is difficult to
say which of the two $\omega$'s to take.

It is interesting to notice that even with a rescaling of the lowest
quasiparticle energy shown in Fig.~\ref{eqp:min}, which roughly may be
identified with the gap, the sharp drop across the drip point is still
there, even though less pronounced. For example multiplying the case
$r=0.1$ (black solid line) by a factor ten beyond $\mu = 26$ will not
reduce the break down by a large amount. Therefore, \emph{e.g.}, the moment
of inertia of a deformed slowly rotating trap would be much less close
to its irrotational flow value after the drip than
before. Even in an almost full finite container this would be the case, since the lowest quasiparticle energy, \emph{i.e.}, implicitly the gap, is strongly reduced already before the top of the container is reached, see Fig. 5.This is an interesting and, may be, unexpected
physical consequence of our study. It should apply equally to other
dynamical quantities as the already mentioned collective modes, see, \emph{e.g.}, Ref~\cite{gra05}.\\

On the other hand, if the system enters the nonlinear regime where vortices are created, certainly, in the steady state configuration, the first appearing vortex will be centered at the origin where the gap is maximum independent of what happens around the Fermi surface.

%
%
\section{Conclusions}\label{sec:concl}
In this work, we enlarged and refined our previous study on superfluid Fermi
systems at overflow. This means having a container which passes as a
function of filling, $i.e.$, as a function of chemical potential
abruptly from a narrow to a wide container. The transition point is
called the overflow or drip point. Our initial interest came from
nuclear physics where such situations may be relevant for very neutron rich 
nuclei in the laboratory or for nuclei in
the  crust of neutron stars. 
There, as a function of the deepness
in the crust, the nuclei may or may not be embedded in a gas of
dripped superfluid neutrons. In this work we studied as a generic
example the situation with cold atoms where a trapping potential of
this kind had already been fabricated but for bosons. However,
theoretical studies have also been performed with fermions
\cite{Viverit2001}. Eventually, one may also arrange superconducting
metallic grains geometrically and energetically in such a way that a
double well with varying chemical potentials is created. In finite
nuclei, the effects at the neutron drip is strongly overshadowed by
shell fluctuations, but a clear average tendency of reduced pairing is
also visible there~\cite{Pastore2013}.

The technical improvement compared to our earlier publications here is
that we treat the cold atom situation with thousands of atoms not in
the BCS approximation (\emph{i.e.}, keeping only the diagonal elements of the
gap) but rather with the BdG~(HFB) approach which is more realistic and 
even absolutely
necessary if one wants to use the standard renormalization method for the
contact interaction  where one replaces the bare interaction by the scattering length.  We, indeed, found that the BCS approximation fails in the case of cold atoms  when a contact force is used together with the usual regularisation scheme.  In addition, the BCS approximation turns out to be inadequate for one of the conclusions advanced in ~\cite{Schuck2011}, namely that the drop of pairing after the drip also entails a corresponding reduction of the gap in the center of the trap. On the contrary, as we show in this work, with the BdG-HFB approach, the gap at the center remains rather unaffected passing through the drip point in a monotonous way. 
On the other hand, we have seen that qualitatively the
main conclusions with respect to our earlier BCS ones remain
unchanged. This concerns, for instance, the finding that the spectral
gaps drop surprisingly sharply across the overflow point and a strong reduction can already be seen before reaching the break
point. In fact, in the limit where the wide container completely opens up, we reach the situation of a single finite potential well. We discussed that in this case, the spectral gap drops to zero when the chemical potential reaches the upper limit, the reduction of the gap starting already well before this limit is reached. We have shown that this effect is also present in  the level density
which in some systems may be a measurable quantity.
We also investigated the wave functions of the lowest quasi-particle states and found that they are concentrated in the surface. This explains the drop of the gap after the drip as demonstrated with the calculation of the lowest quasi particle energy and, also, the  level density before and after the overflow.\\

In Refs.\cite{Schuck2011,Schuck2012}, we
also studied the Thomas-Fermi (TF) version of BCS using a cut off or a finite range interaction and obtained very
good agreement with the corresponding quantal BCS results. An extension of
TF to HFB or BdG is planned for the future. 

 It may be noticed that for cold atom systems, we give a definite experimental prediction in this work. We also discussed
the influence of our finding on dynamic quantities like moment of
inertia and other flow aspects.


\acknowledgments

One of us (X.V.) acknowledges financial support from Spanish Consolider-Ingenio 2010 Programme CPAN CSD2007-00042
and from Grants No.FIS2011-24154 MICINN and FEDER and No.2009SGR-1289 from Generalitat de Catalunya.
We thank A. Polls for interesting discussions on recent theoretical progress in cold atom physics.


\bibliography{biblio}

\end{document}